\documentclass[journal=langd5,manuscript=article,layout=twocolumn]{achemso} 
\usepackage{chemformula} % Formula subscripts using \ch{}
\usepackage{amssymb}
\usepackage[T1]{fontenc} % Use modern font encodings
\usepackage[switch, modulo]{lineno}
\author{Hrishikesh Pingulkar}
\affiliation[LOMC]{LOMC, CNRS and Universit\'{e} Le Havre Normandie, Le Havre, France}
\author{Jorge Peixinho}
\affiliation[Second University]{Laboratoire PIMM, CNRS, Arts et M{\'{e}}tiers Institute of Technologie, Cnam, Paris, France}
\alsoaffiliation[LOMC]{LOMC, CNRS and Universit\'{e} Le Havre Normandie, Le Havre, France}
\email{jorge.peixinho@ensam.eu}
\author{Olivier Crumeyrolle}
\affiliation[LOMC]{LOMC, CNRS and Universit\'{e} Le Havre Normandie, Le Havre, France}

\title[Liquid Transfer for Viscoelastic Solutions]{Liquid Transfer for Viscoelastic Solutions} 

\begin{document}

%% The "tocentry" environment can be used to create an entry for the graphical table of contents. It is given here as some journals require that it is printed as part of the abstract page. It will be automatically moved as appropriate.

%\begin{tocentry}
%\includegraphics[width=1.0\columnwidth, keepaspectratio]{Figures/TOC1.eps}
%Some journals require a graphical entry for the Table of Contents.
%This should be laid out ``print ready'' so that the sizing of the text is correct.

%Inside the \texttt{tocentry} environment, the font used is Helvetica 8\,pt, as required by \emph{Journal of the American Chemical Society}.

%The surrounding frame is 9\,cm by 3.5\,cm, which is the maximum permitted for \emph{Journal of the American Chemical Society} graphical table of content entries. The box will not resize if the content is too big: instead it will overflow the edge of the box.

%This box and the associated title will always be printed on a separate page at the end of the document.
%\end{tocentry}

\begin{abstract}
Viscoelastic liquid transfer from one surface to another is a process that finds applications in many technologies, primarily in printing. 
Here, cylindrical shaped capillary bridges pinned between two parallel disks are considered. 
Specifically, the effects of polymer mass fraction, solution viscosity, disk diameter, initial aspect ratio, final aspect ratio, stretching velocity and filling fraction (alike contact angle) are experimentally investigated in uniaxial extensional flow. 
Both Newtonian and viscoelastic polymer solutions are prepared using polyethylene glycol (PEG) and polyethylene oxide (PEO), with a wide variety of mass fractions.
The results show that the increase in polymer mass fraction and solvent viscosity reduces the liquid transfer to the top surface.
Moreover, the increase in the initial and final stretching height of the capillary bridge also decreases the liquid transfer, for both Newtonian and viscoelastic solutions.
Finally, the shape of the capillary bridge is varied by changing the liquid volume.
Now, Newtonian and viscoelastic solutions exhibit opposite behaviors for the liquid transfer.
These findings are discussed in terms of interfacial shape instability and gravitational drainage.
\end{abstract}

\section{Introduction}

Liquid capillary bridges can be found in many industrial applications such as food processing, material engineering, adhesion processes, coating technology, flow in porous media, microfluidics and measurement of rheological properties \cite{Kumar2015,Gutierrez2020,karim2020,Montanero2020}.
One of the motivations for the study of stretched liquid bridges is its close association with printing processes.
Nowadays, a printing process is not only limited to books, magazines and newspapers but has also expanded to various areas like manufacturing of electric circuits \cite{Kumar2015}, printed wearable electronics \cite{Sekine2018}, screen displays, lab-on-a-chip \cite{Park2007}, solar cells \cite{Krebs2009} and 3D microstructures (polymer wires, needles, pillars, cones, and microspheres) \cite{Grilli2011}.
The printing industry deals with inks, which can contain polymers, surfactants or particles, and have viscoelastic properties. 
While printing, the liquid from one surface is transferred to another surface through the formation of a capillary liquid bridge. 
For Newtonian fluids, during this liquid transfer, solution pools are formed on the end plates of a stretched liquid bridge.
On the other hand, a stretched viscoelastic liquid bridge forms a persistent thin filament \cite{Goldin1969,Oliveira2006,Tirtaatmadja2006,Dinic2017,Ponce-Torres2017,Pingulkar2020} along with the solution pools.

\begin{figure}[tb]
\centering
\includegraphics[width=1.0\columnwidth]{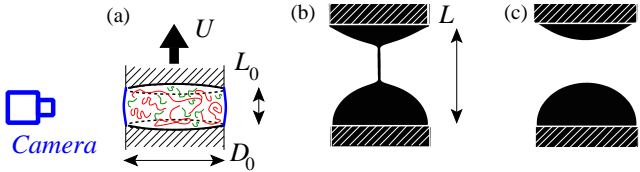}
\caption{(a) Liquid capillary bridge of polymer solution sandwiched between parallel disks of diameter, $D_0$, and initial height, $L_0$, before being stretched with a velocity, $U$. 
(b) The top disk reaches the final stretching height, $L$, a thinning filament is observed.
(c) Breakup of the liquid bridge.}
\label{fig1}
\end{figure}

Due to complexities involving surface and liquid properties, as well as the formation of filaments, the transfer of desired volume of liquid is a challenge in the printing industry.
Specifically, capillary, viscous, inertial, elastic, and gravitational forces all play a role in this liquid transfer.
A relevant parameter called the transfer ratio, $T_r$, is defined as the fraction of liquid transferred to the moving disk (accepter), to the total amount of liquid left on both disks, as shown in Figure \ref{fig1}.
The geometrical properties that affect $T_r$ are the disk radius, $R_0$, or the disk diameter, $D_0$, the stretching speed, $U$, the initial aspect ratio, $L_0/R_0$, and the final aspect ratio, $L/R_0$. 

\citeauthor{Chadov1979} \cite{Chadov1979,Yakhnin1983} first identified the liquid transfer phenomenon for several liquids of viscosity, $\eta$, and surface tension, $\sigma$, on various surfaces.
In their experiments, the top flat surface was brought downwards to press a liquid drop on the bottom surface, and then moved upwards to stretch the capillary bridge at high capillary numbers, i.e. $Ca=\eta U / \sigma >1$.
In these conditions, the liquid transfer is independent of surface and liquid properties, which was later confirmed by others authors\cite{Kang2008,Ahmed2011,Chen2015,Huang2016}. 
Noteworthy, when the contact angles on the both plates are 45$^\circ$, $T_r$ is found to be 0.5.

The issues of wetting and the dynamic contact angle\cite{DeGennes1985} are generally avoided by pinning the contact line to the edge of a disk.
For example, \citeauthor{Zhang1996}\cite{Zhang1996} investigated, theoretically and experimentally, the stability and the breakup of the liquid bridge, with fixed contact lines at the edges.
The authors found $T_r$ decreases with increasing disk diameter, but increases with the stretching speed.
In practice, gravure cells are used \cite{Dodds2009,Dodds2011,Sankaran2012,Lee2013} to control the ink spreading.
The effect of viscoelastic polymer solutions on $T_r$ for the gravure printing is studied experimentally by \citeauthor{Sankaran2012} \cite{Sankaran2012}, and numerically by \citeauthor{Lee2013} \cite{Lee2013}
In the beginning, the liquid transfer is governed by the early stretching dynamics, while the final amount of liquid transfer occurs through the delayed viscoelastic filament thinning \cite{Sankaran2012,Lee2013,Wu2019}.
In contrast, the present study focuses on the cylindrical shaped liquid bridges, with the pinned contact lines on disks to overcome the shearing effect \cite{Wu2019}.

In the present paper, our experimental study mainly focuses on $T_r$ for a cylindrical-shaped bridge of viscoelastic, as well as Newtonian fluids.
Solutions will be prepared by varying mass fraction and molecular weight of polymers to investigate the effect of viscoelasticity on liquid transfer. 
The same stainless steel disks at the top and bottom will be used to overcome the complexities due to different surface materials as previously reported \cite{Chadov1979,Yakhnin1983,Wu2019}.
Different disk diameters, initial and final aspect ratios, along with the initial profile curvatures are explored.

\section{Materials and methods}

\subsection{Sample preparation}

Aqueous solutions of poly-ethylene oxide (PEO), which is a high molecular weight polymer, and poly-ethylene glycol (PEG), which is a relatively low molecular weight, are used, either separately or in combination. 
PEO solutions are commonly used to investigate viscoelastic behavior of liquid jets \cite{Christanti2001,Tirel2017}, beads-on-a-string \cite{Rodd2005,Pingulkar2020,Oliveira2006,Deblais2018}, coatings \cite{Bazzi2019,Gaillard2019,Karim2018} and liquid transfer \cite{Sankaran2012,Wu2019}.
Broadly, our strategy is to control the shear viscosity with PEG and the extensional viscosity with the mass fraction of PEO. 
The molecular weight of PEO is $8\times10^6$ g/mol and the molecular weight of PEG is 20 000 g/mol, according to Sigma-Aldrich. 
Three different types of solutions are prepared: (i) aqueous PEG solutions, (ii) aqueous PEO solutions and (iii) mixtures of PEG and PEO solutions.
For the solutions containing PEO, 0.5 wt.\% iso-propyl alchohol is added for easy dispersion of PEO molecules in solvent\cite{Layec1983}.
The mass fraction of PEO in solutions ranges from 100 to 2000 ppm with and without PEG. 
Note, a solution of 1000 ppm of PEO is labelled PEO1000.
Similarly, a solution of 20 wt.\% PEG and 2000 ppm of PEO is labelled PEG20PEO2000.
The density, $\rho$, and $\sigma$ were measured for each solution and the values are reported in Table 1 in the Supporting Information. 

\subsection{Rheological measurements}

\begin{figure*}[t]
\includegraphics[width=2\columnwidth]{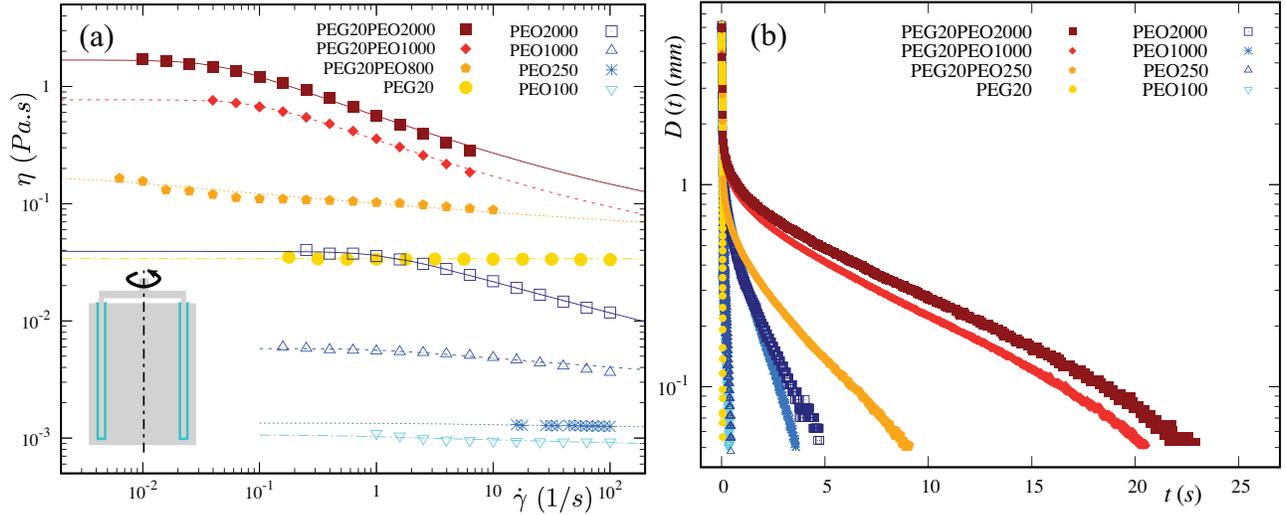}
\caption{(a) Steady shear viscosity against shear rate and (b) diameter time evolution for aqueous solutions of PEO, PEG and PEG20PEO. The inset in (a) is a sketch of the double wall geometry (drawn up to scale) used and the lines represent fits described in the text.}
\label{fig2}
\end{figure*}

The shear viscosity of the solutions was measured using rotating rheometer (TA Instruments Discovery HR-3) with double wall concentric cylinder geometry. 
High polymer mass fraction solutions exhibit shear thinning, as shown in Figure \ref{fig2}(a).
The zero shear viscosity, $\eta_0$, was obtained from fitting the Carreau equation \citep{Bird1987} with the infinite viscosity prescribed to the solvent one.

% Comment Fig. 2(a-b)
Additional rheological measurements were conducted using a capillary breakup extensional rheometer (CaBER from Thermo Haake).
The filament diameter around mid-height is tracked over the time using an in-built laser (see Figure \ref{fig2}(b)).
The diameter decreases with time for all solutions. 
For viscoelastic solutions, the diameter exhibits an exponential thinning, characterised by the extensional relaxation time, $\lambda$.
Further analysis of the diameter can be used to obtain the extensional viscosity as a function of time and Hencky strain.
These results are given in \citeauthor{Pingulkar2020}\cite{Pingulkar2020}

Based on the solution properties, several dimensionless numbers can be defined.
The Deborah number, $De=\lambda/\tau_R$, represents a ratio of the viscoelastic time, $\lambda$, to the Rayleigh inertio-capillary time: $\tau_R=\sqrt{\rho {R_0}^3/\sigma}$.
Other authors \cite{Bhat2010,Lee2013} use the ratio $\lambda/\tau_v$ with the viscous time $\tau_v=\eta_0 R_0/\sigma$.
In addition, the ratio of the viscous and Rayleigh times is also needed, that is the Ohnesorge number: $Oh=\tau_v/\tau_R$. 
Another important number to characterize a capillary bridge is the Bond number quantifying the effect of gravitational to capillary forces: $Bo=\rho g{R_0}^2/\sigma$. 
Later, when the capillary bridge shape is non-cylindrical, the mid-height radius, $R_{mid}$, will be used, leading to $Bo_m$.
The experiments use three disk diameters, additional mass fractions, and parameters are presented in Table 2 in the Supporting Information.

\subsection{Experimental setup}

The capillary bridge stretching is studied experimentally with the CaBER and a high-speed camera (800$\times$1280 pixels).
The resolution for typical experiments is 1 pixel = 2 $\mu$m.
Shadowgraphic images are obtained using a continuous laser and a diffuser (from Dantec dynamics).
Initially, a liquid capillary bridge is created by placing a solution sample in-between the two parallel disks.
The volume of the solution introduced in the liquid bridge depends on $R_0$ and the initial height, $L_0$, and this volume is calculated as $\mathcal{V}=\pi L_0 R^2_0$.
The liquid volume introduced is controlled with the help of a pipette (Eppendorf research plus).
After stretching of the liquid bridge, solution pools are formed on the top and bottom disks, as shown in the binarized image in Figure \ref{fig1}(c).
Considering the axisymmetric capillary bridge, images of the vertical cross-section of the solution pools are captured.
Then, heights, $h_T$ and $h_B$, of top and bottom solution pools, respectively, are estimated by counting the pixels in the appropriate directions.
The truncated sphere method \cite{Kang2008} is implemented to calculate the volume of the solution pools: 
$V_{T,B} = \pi h_{T,B} \left( 3 R^2_0 + h^2_{T,B} \right) / 6$.
So, the transfer ratio, $T_r$, is defined in the same way as previous authors\cite{Yakhnin1983,Kang2008,Chen2014,Huang2016,Tourtit2019} as: $T_r = V_T/(V_T + V_B)$, 
In the following, every $T_r$ data point and the associated error bar symbolise the average of five experiments and their dispersion, respectively,

\section{Experimental results and discussion}

The results consist of a series of experiments that report the $T_r$ for different polymer mass fractions, disk diameters, final heights and, finally, different shapes of the initial capillary bridge. 
The resulting $T_r$ are discussed in terms of morphology of interface and gravitational drainage.

\subsection{Effect of polymer mass fraction}

\begin{figure*}[thb]
\centering
\includegraphics[width=2.0\columnwidth]{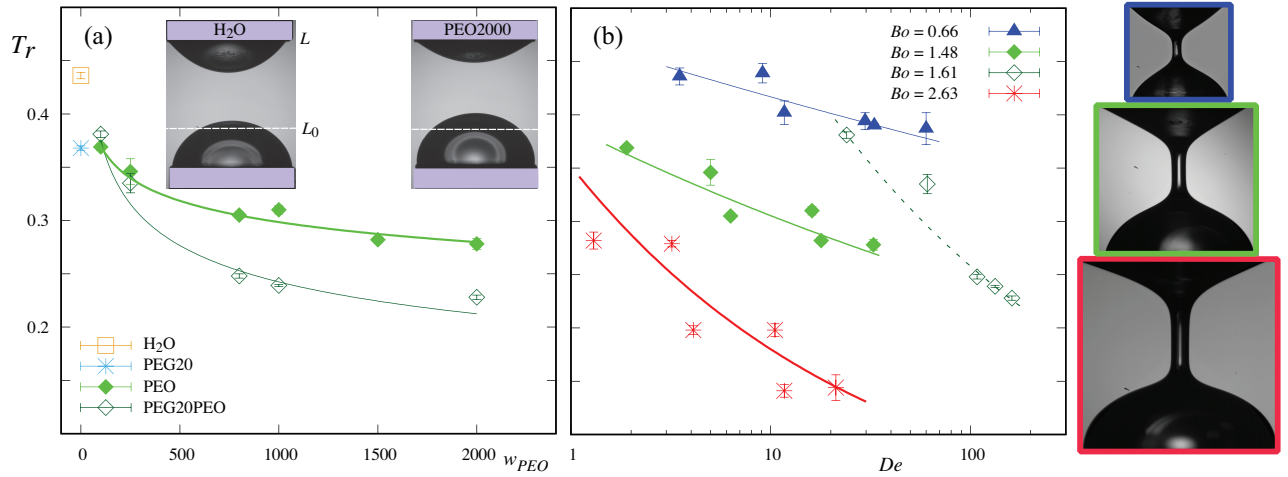}
\caption{(a) Transfer ratio, $T_r$, as a function of the polymer mass fraction, $w_{PEO}$.
$R_0=3$ mm, $L_0/R_0=0.66$ and $L/R_0=2$ at $U=0.08$ m/s.
(b) Transfer ratio, $T_r$, as a function of $De$ for different $Bo$. 
The insets are stretched liquid bridges for 1000 ppm of PEO for $R_0=2$, 3 and 4 mm, respectively (from top to bottom).}
\label{fig3}
\end{figure*}

The PEO mass fraction is varied, in the aqueous PEO and PEG+PEO solutions, to test its influence on $T_r$.
Experiments are preformed using water, PEG20, PEO and PEG+PEO solutions.
For these experiments, cylindrical shaped initial liquid bridges are prepared with $R_0=3$ mm and $L_0/R_0=0.66$.
Then, the liquid bridges are stretched to $L/R_0=2$ at $U=0.08$ m/s.
These initial capillary bridge are inside the stability limit curve \cite{Bezdenejnykh1992}.
The $T_r$ obtained are presented against the PEO mass fraction, $w_{PEO}$, in Figure \ref{fig3}(a).
For both the PEO and PEG+PEO solutions, $T_r$ decreases with increasing $w_{PEO}$.
The same can be confirmed from the inset photographs where a smaller top solution pool for PEO2000 solution compared to water can be observed.

Typically, for a viscoelastic liquid bridge, the liquid transfer occurs in two stages.
In the first stage, when a liquid bridge is stretched, the minimum radius location along the filament appears and an initial liquid transfer takes place from the bottom to the top disk.
When $Bo>1$ for $R_0=3$ mm (see Table 2 in the Supporting Information), an initial asymmetry (sagging) decides the primary liquid distribution along the stretched liquid bridge.
Hence, when the liquid bridge is stretched, initially a larger bottom solution pool is formed compared to the top solution pool.
However, for PEO and PEG+PEO solutions, solution pools are formed along with a filament that plays a significant role in the second stage of liquid transfer.
In this second stage, liquid is transferred from the top to the  bottom solution pool due to gravitational drainage through the filament.
As observed in Figure \ref{fig2}(b), with increasing $w_{PEO}$, the filament lasts longer and liquid bridge rupture gets delayed\cite{Bazzi2019}.
As a result, there is more time for the gravitational drainage and more liquid is transferred from the top to the bottom solution pool.
For low mass fraction, for both PEO and PEG+PEO solutions, the filament breakups early and there is less time available for the liquid transfer from the top to bottom solution pool.
Hence, the viscoelastic liquids nearly reproduce the response of PEG20, as reported by \citeauthor{Sankaran2012} \cite{Sankaran2012}.
Furthermore, for PEG+PEO solutions, filament lasts longer compared to the PEO solutions.
Hence, because of the higher gravitational drainage, lower values of $T_r$ are obtained for the PEG+PEO solutions compared to the PEO solutions.
Power law fits are used to capture the change in $T_r$ with $w_{PEO}$ and the exponents for PEO and PEG+PEO solutions are 0.09 and 0.19, respectively.
The larger exponent of the fit for the PEG+PEO solutions than for the PEO solutions suggests that effect of change in $w_{PEO}$ is higher for the PEG+PEO solutions.

Furthermore, for PEO and PEG+PEO solutions, $T_r$ is plotted as a function of $De$ in Figure \ref{fig3}(b).
For both solutions, $T_r$ decreases with increasing $De$.
Similar behavior was observed numerically by \citeauthor{Lee2013} \cite{Lee2013}, where the authors reported that liquid transferred to the top disk decreases with increase in $De$.
However, these results were obtained for a combination of gravure cell at the bottom and a flat plate moving upwards.
Again, power law fits can be used to represent $T_r$ and the exponents differ for PEG+PEO solutions (-0.27 for $Bo=1.61$) than PEO solutions (-0.10 for $Bo=1.48$).

\subsection{Effect of disk radius}

% Comments on Fig. 3(b)
One of the important parameter of the capillary bridge geometry is the disk radius. 
To understand the influence of $R_0$ on $T_r$, disks of different radii, such as 2, 3 and 4 mm, are tested, by keeping other geometric parameters the same.
As a result, depending on $R_0$, the Bond number, $Bo$, varies from 0.66 to 2.63.
For these set of experiments, $L_0/R_0$ and $L/R_0$ are kept constant at 0.66 and 2, whereas, the stretching speed, $U$, is again 0.08 m/s. 
It can be observed that for all disks radii tested, $T_r$ decreases with $De$.
Additionally, the increase in $R_0$ as well as $Bo$ globally leads to the decrease in $T_r$, and indicates the effect of gravity.
A similar decrease in $T_r$ with $R_0$ has been reported for water by \citeauthor{Zhang1996} \cite{Zhang1996}

In the photographs of Figure \ref{fig3}(b), the stretched liquid bridges with different $R_0$ are displayed for PEO1000 solution at $L/R_0 = 2$.
With increase in $R_0$ from 2 to 4 mm, the shape of the top solution pools weakly changes from convex to concave in the vertical plane, but the shape of the bottom solutions pool remains convex.
However, the capillary pressure  associated to the shape change is dominated by hydrostatic pressure difference ($\rho g L$).
In addition, for the largest $R_0$, the hydrostatic pressure difference overcomes the capillary pressure associated to the mean curvature in the horizontal planes of the top solution pool.
Hence, with increasing $R_0$, more solution is pushed from the top solution pool towards the bottom solution pool due to increased stretched liquid bridge height (fixed $L/R_0=2$). 
Furthermore, it has been observed that the filament thinning time increases with increasing $R_0$.
Hence, the gravitational drainage through the filament is enhanced by the larger disk radius that results in lower values of $T_r$.
For each $Bo$, power law fits are again used to capture the change in $T_r$ with $De$.
With increasing $Bo$, the power law exponents increase from 0.05 to 0.29, suggesting larger effect of $De$ on higher $Bo$ due to combined effect of gravity and delayed filament thinning.
For each $w_{PEO}$, $T_r$ decreases linearly with $Bo$. 
A reduced plot is shown in Supplementary Information (Figure S1(b)). 

\begin{figure}[htb]
\centering
\includegraphics[width=1\columnwidth, keepaspectratio]{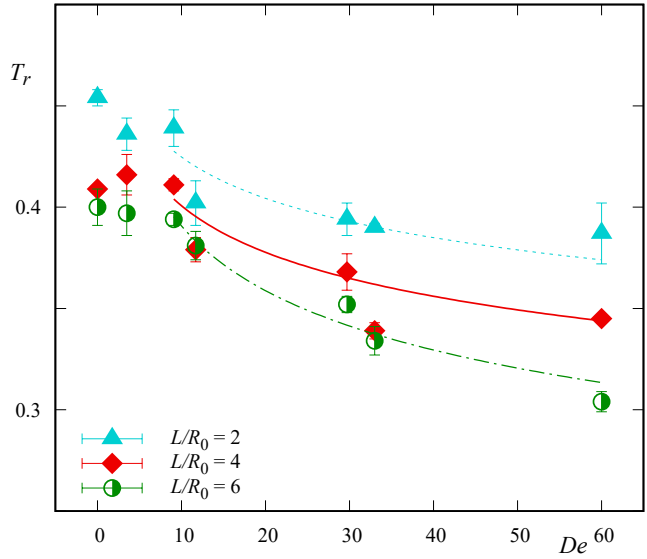}
\caption{$T_r$ as a function of $De$ for different $L/R_0$ using PEO solutions.
The liquid bridge, formed in-between the two disks having $R_0=2$ mm is stretched from ${L_0}/R_0=0.66$ to different $L/R_0$ at $U=0.08$ m/s.}
\label{fig4}
\end{figure}

\subsection{Effect of final stretching height}

Effect of the final stretching height on the liquid transfer is explored by varying the final aspect ratio, $L/R_0$, from 2 to 6.
All other geometrical and stretching parameters are kept constant.
The minimum value of $L/R_0$ is limited by the least value of the stretching height required for breaking of the capillary bridge, whereas the maximum value of $L/R_0$ is limited by the experimental constraints.
Two different experimental studies are carried out for the PEG20 and PEO solutions.

\paragraph*{PEG20 solution}

The obtained results for the variation of $T_r$ are plotted, against $L/R_0$, in Figure S3 in the Supporting Information.
The photographs illustrate different shapes of the capillary bridges formed, just before their breakup, when the top disk is at $L/R_0=2$, 4, and 6, respectively.
The $T_r$ decreases linearly with increase in the $L/R_0$.
For $L/R_0=6$, the breakup occurs close to the top solution pool, compared to the breakup at the middle of the bridge, for $L/R_0=2$.
This further suggests that, after breakup, the liquid transferred to both disks will be nearly the same for $L/R_0=2$.
But, with increasing $L/R_0$, as the breakup point moves closer to the top solution pool, more liquid volume will be enclosed below this breakup point and a larger solution pool is formed on the bottom disk.
Hence, with increasing final stretching height, less liquid will be transferred to the top disk.

\paragraph*{PEO solutions}

The change in $T_r$, for  values of $L/R_0=2$, 4 and 6 are plotted against $De$, as shown in Figure \ref{fig4}.
For all $L/R_0$, with the increase in $De$, $T_r$ decreases.
Additionally, for the same $De$, it can be noted that the increase in $L/R_0$ leads to the decrease in $T_r$.
This behaviour appears to be more significant for $De>10$.
Hence, for $De>10$, power law fits of $T_r$ are used for each $L/R_0$.
With increasing $L/R_0$, the value of the power law exponent increases from 0.07 to 0.12.
The location of the necking points, on the liquid bridge, varies with different $L/R_0$, in a similar manner to PEG20 (see photographs in Figure S3 in the Supporting Information).
Length of the filament increases with increasing $L/R_0$ and hence, larger liquid volume is enclosed below this necking point.
Again, the filament lasts longer with the increase in $De$, which further helps for the drainage.
This combined effect of location of necking point and drainage produces the smaller values of $T_r$ at higher $L/R_0$ and $De$.

Effects of stretching speed on liquid transfer are discussed in details in the  Supporting Information in Figure S4.
For PEO solutions our results show, that $T_r$ does not increase significantly with increasing $U$,  due to $Ca \le 0.01$.

\subsection{Effect of initial bridge shape}

\begin{figure*}
\centering
\includegraphics[width=2.0\columnwidth, keepaspectratio]{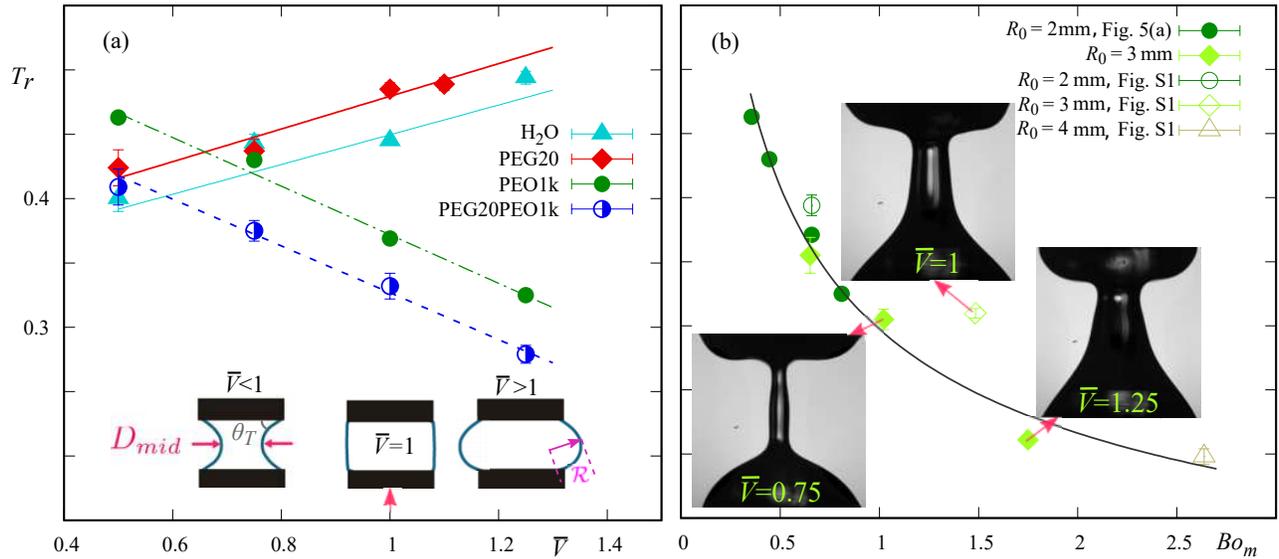}
\caption{(a) $T_r$ as a function of $\overline{V}$ for Newtonian and viscoelastic solutions.
Liquid bridges, formed in-between two disks having $R_0=2$ mm are stretched from $L_0/R_0=0.66$ to $L/R_0=2$, at $U=0.134$ m/s.
The liquid volume introduced is varied from 8 to 20 $\mu$l ($0.5<\overline{V}<1.25$), as illustrated in the inset.
(b) $T_r$ as a function of $Bo_{m}$ for PEO1000 over a range of $R_0$ and $\overline{V}$.
The trend line is a power law fit with $T_r\propto Bo_m^{-0.46}$.
The photographs depict stretched liquid bridges for various $\overline{V}$ with $R_0=3$ mm, when the top disk reaches $L/R_0=2$, with arrows indicating the associated data points.
$\overline{V}=1$, for the points from Figure S1.
$U=0.08$ m/s, except for data from Figure 5(a).}
\label{fig5}
\end{figure*}

The influence of initial bridge shape on liquid transfer is studied by varying the liquid volume introduced with the pipette.
This also plays a role in setting the initial liquid contact angles.
For a cylindrical shaped liquid bridge, liquid volume is $\mathcal{V}$.
Then, a dimensionless parameter, $\overline{V}$, is defined as a ratio of the actual liquid volume introduced to $\mathcal{V}$.
The cylindrical-shaped liquid bridge, for which $\overline{V}=1$, and the associated contact angle on the top disk, $\theta_T \simeq 90^{\circ}$, are illustrated in the inset of Figure \ref{fig5}(a).
$\overline{V}$ is varied from 0.5 to 1.25 resulting in $55^{\circ} \lesssim \theta_T \lesssim 115^{\circ}$.
In our experiments, for $\overline{V}<0.5$ or $\overline{V}>1.25$, the bridge is found to be unfeasible or unstable.

Four different types of fluids, such as water, PEG20, PEO1000 and PEG20PEO1000, are used to characterise the effect of the initial bridge shape on $T_r$.
The liquid bridges, formed with ${R_0}=2$ mm, are stretched, from ${L_0}/R=0.66$ to $L/R=2$, at $U=0.134$ m/s.
The initial bridge shape with the contact angles, $\theta_T< 90^{\circ}$, at the top disk is similar to previous works \cite{Kang2008,Wu2019,Ahmed2011,Chen2014}, where slipping contact lines were observed.
Yet, in our case, the liquid contact lines are pinned, for all initial bridge shapes.
%, possibly because of the pre-wetting of the disk surface
The aqueous solutions have the tendency to wet the hydrophilic stainless steel disks. 
Clearly PTFE disks would lead to unpinning and sliding of the contact lines, which in turn induce an unwanted interface dynamics\cite{Kumar2015,Wu2019} in the present study.
The obtained $T_r$ results are plotted against the increasing $\overline{V}$, in Figure \ref{fig5}(a). 
It can be observed that $T_r$ increases with increase in $\overline{V}$ for water and PEG20. 
But, for PEO1000 and PEG20PEO1000 viscoelastic solutions, the increase in $\overline{V}$ has exactly opposite effect, where $T_r$ decreases with increase in $\overline{V}$.
%In addition, with increase in $\overline{V}$, this opposite effect on liquid transfer increases.
%The linear fits are used to compare the variation of $T_r$ with $\overline{V}$. 
%The slopes of the fit for PEG20 and water are identical, whereas the slopes for PEO1000 and PEG20PEO1000 are the same.
%The higher values of the slopes for PEO1000 and PEG20PEO1000 solutions, than water and PEG20, suggest that the change in volume, and hence, the contact angles, have stronger effect on the $T_r$ for the viscoelastic solutions than for the Newtonian fluids.
%The additional plots are given in Fig. 2 of the supplementary material, for the effect of initial bridge shape, for disk diameter, ${D_0}=6$ mm.
%The similar trends are observed for PEG20 and PEO1000 solutions, with ${D_0}=6$ mm, where the $T_r$ decreases with increase in \% $V$, for the viscoelastic solutions, and the $T_r$ increases with increase in \% V, for the Newtonian solutions.

The effect of the initial bridge shape is further investigated, for PEO1000, by measuring the mid-plane diameter, $D_{mid}$, of the initial liquid bridge, for $R_0=2$ and 3 mm, for all $\overline{V}$, using image analysis.
The inset photographs illustrate different shapes of PEO1000 liquid bridges formed for various $\overline{V}$ with $R_0=3$ mm when the top disk is at $L/R_0=2$.
With increase in $\overline{V}$, larger filament radius can be observed that enhances the gravitational drainage through the filament.
Note that this filament radius after stretching is related to $R_{mid}$.
Then, $T_r$ is plotted in Figure \ref{fig5}(b) against the modified Bond number calculated using $R_{mid}$, that is $Bo_m=\rho g R_{mid}^2/\sigma$ (see inset of Figure \ref{fig5}(a) for a sketch with $D_{mid}$). 
Other definitions of the Bond number have been proposed\cite{Mazzone1986,Willett2000}.
Here, a power law fit is obtained with $T_r\propto Bo_m^{-0.46}$ and a coefficient of determination of 0.92.
This results in $T_r$ being proportional to $R_{mid}^{-0.92}$.
For $\overline{V}=1$, it was reported that $T_r$ decreases linearly with $Bo$ (see Figure S1(b) in the Supplementary Information). 
However when $\overline{V}\neq 1$, the curvatures controlling the Laplace pressures are changed while the height, $L_0$, and thus hydrostatic pressure, is not. 
Hence a physics based formulation for the Bond number would be $$Bo^*=\frac{\rho g L_0}{\sigma/R_{mid}+\sigma/\mathcal{R}}$$ $\mathcal{R}$ is the radius in the vertical plane (see sketch in the inset of Figure \ref{fig5}(a)), axisymmetric as a first approximation. 
As $\sigma/\mathcal{R}$ changes sign when $\overline{V}$ varies around one, $Bo^* \simeq \rho g L_0 R_{mid}/\sigma$.
Our results on $Bo_m$ converts into $T_r$ being close to a linear decrease with $Bo^*$, rather than $Bo$.

\section{Conclusions}

Liquid transfer, for viscoelastic polymer solutions, was studied experimentally. 
Reference results using Newtonian liquids on the effect of the initial and final aspect ratio have shown a linear decrease of transfer ratio.
Then, the cylindrical shaped liquid capillary bridge, with pinned contact lines between two parallel disks, was stretched for a range of polymer mass fractions.
The transfer ratio decreases with Deborah number and can be explained by the gravitational draining enhanced by the delayed filament thinning.
The gravitational influence was further studied by using different disk radii and the results show that transfer ratio decreases for viscoelastic fluids with increasing disk radius, similar to Newtonian fluids.
From the printing point of view, smaller values of polymer mass fraction, disk radius, as well as initial stretching height along with final stretching height should be favored for larger liquid transfer.
For non-cylindrical shaped bridges, the liquid transfer behavior vary sensitively with the liquid volume introduced.
Moreover, the behavior of the transfer for Newtonian solutions increases as a function of actual liquid volume introduced and decreases with viscoelastic solutions.
Then, the trend for results with PEO1000 follow a power law with $Bo_m^{-0.46}$, where $Bo_m$ is the Bond number computed from the initial mid-plane radius. 
Another Bond number, $Bo^*$, with explicit curvatures could be used, but can have negative values. 
Hence, for future studies on transfer ratio, the experimentally tractable $Bo_m$, based on actual radius $R_{mid}$ is suggested.\\

\begin{acknowledgement}
%Please use ``The authors thank \ldots'' rather than ``The authors would like to thank \ldots''.
The authors thank  M.-C. Renoult for discussions on the downward stretching and M. Grisel for access to a rheometer with double wall concentric cylinders. 
Partial funding for this research was provided by the project BIOENGINE, which was co-financed by the European Union with the European Regional Development Fund and the R\'{e}gion Normandie, the  French Agence Nationale de la Recherche and LabEx EMC3 through the project IBOASD (Grant No. ANR-10-LABX-09-01) and the Universit\'{e} Le Havre Normandie (support of H. Pingulkar).
\end{acknowledgement}

\begin{suppinfo}
Details of the fluid properties, including dimensionless numbers; additional results on $T_r$ for (i) the effect of $R_0$ and $Bo$, (ii) the effect of the initial aspect ratio: $L_0/R_0$, (iii) the effect of the final aspect ratio: $L/R_0$ and (iv) the effect of the stretching velocity.
This material is available free of charge via the Internet at http://pubs.acs.org.
\end{suppinfo}

\bibliography{NewVisco} 

\end{document}